\documentclass[twocolumn]{revtex4}

\usepackage{epsfig}
\usepackage{amssymb}
\usepackage{amsmath}
\usepackage{amsfonts}
\usepackage{epstopdf}
\usepackage{graphicx}
\usepackage{bm}
\begin{document}

\title{Temperature dependent nonlinear Hall effect in macroscopic Si-MOS antidot array.}

\author{A.\,Yu.\,~Kuntsevich$^{+*}$,
A.\,V.\,Shupltetsov$^{+*}$, M.\,S.\,Nunuparov$^{o}$\/\thanks{e-mail: alexkun@lebedev.ru}}

\address{$^{+}$ P.N. Lebedev Physical Institute of the RAS, 119991 Moscow, Russia \\
          $^{*}$ Moscow Institute of Physics and Technology, 141700 Moscow, Russia\\
          $^{o}$ Prokhorov General Physics Institute, Russian Academy of Sciences, 119991, Moscow, Russia\\
          }

\begin{abstract}
By measuring  magnetoresistance and Hall effect in classically moderate perpendicular magnetic field in Si-MOSFET-type macroscopic antidot array we found a novel effect: nonlinear with field, temperature- and density-dependent Hall resistivity. We discuss qualitative explanation of the phenomenon and suggest that it might originate from strong temperature dependence of the resistivity and mobility in the shells of the antidots.
\end{abstract}

\maketitle

Low-field Hall resistance ($R_{xy}$) is broadly used to determine electron density. In two dimensions one has $R_{xy}=B/ne$ (where $n$- is electron density, $B$- magnetic field, perpendicular to 2D plane, $e$- electron charge). Remarkably, within the classical treatment, Hall  resistance  coincides with Hall resistivity and does not depend on geometry, e.g. if the hole is drilled in the 2D gas, the current flow will be redistributed while the $R_{xy}$ value will stay the same. On the other hand, such geometrical constriction effectively admixtures Corbino geometry to ordinary Hall bar thus leading to  huge positive magnetoresistance\cite{solin}.

One of the ways to tune experimentally parameters of the drilled holes and 2D gas independently is double-gated antidot array(AA), where electron density at the dots and residual 2D gas is controlled by two independent gate electrodes.
Antidot arrays give a wast parametrical space (mean free path, densities of the 2D gas and antidots, sizes of the antidots, distances between them), transport properties   magnetic field, temperature etc ). As a rule either quantum interference effects(Altshuler-Aharonov-Spivak oscillations\cite{aas}) or quantization effects (commensurability oscillations\cite{weiss}). The global aim of commensurability magnetooscillation studies was to experimentally explore fractal structure of Landau levels, co-called Hofstadter butterfly. This aim was not achieved in semiconducting artificial periodic systems, and experimental progress was achieved only recently in graphene-based natural perfectly periodic system\cite{hofs}. Low-field Hall effect in AA was not addressed that thoroughly, as far as we know.

\begin{figure}
\vspace{0.1 in}
\centerline{\psfig{figure=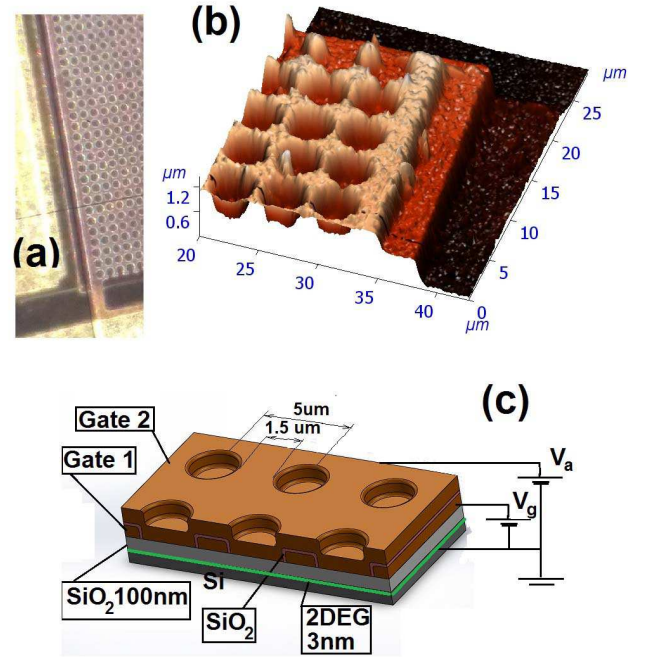,width=250pt}}
\begin{minipage}{3.2in}
\caption{(a) Optical image (b)AFM image and (c)schematics of the gating of the studied antidot array.}
\label{sample}
\end{minipage}
\vspace{0.4 in}
\end{figure}

In order to suppress above mentioned oscillatory effects we study classically large antidots, i.e. both size and distance between the antidots are much larger than mean free path and magnetic length for reasonable fields (few Tesla).

Theoretically, the problem of Hall effect in two-dimensional inhomogeneous system is a very long-standing one (see e.g. \cite{isich} for review). The most recent theoretical calculations were performed by Bulgadaev and Kusmartsev within effective media approach\cite{bulgad1,bulgad2} and by Parish and Littlewood with within random lattice method\cite{litle2}. Both approaches within non-realistic toy models open possibility for the non-linear field dependence of the Hall resistance.

{\bf Samples used:}
We used lithographically defined Hall-bar shaped Si-MOSFETs containing AA with two independent gates: one for antidot array $V_a$ and one for remaining 2D gas (R2DEG) $V_g$. Independently on the same chip there were defined separate Hall bars without antidots obtained within the same processes. We therefore could measure transport properties of both pristine 2D gas and R2DEG independently. The samples were produced similarly to those used in Ref.\cite{reznikov}. The substrate was doped and preserved certain conductivity ($R\lesssim 1$GOhm) downto 20 K, therefore all measurements were performed at lower temperatures.
The dimensions of the Hall bars with AA were 0.4 mm x 0.4 mm;  the dimensions of the Hall bars with pristine 2DEG were 0.1mm x1mm.
All distances and schematical geometry of the antidots are shown in Fig. \ref{sample}.
About 400 chips were defined lithographically on one 4" wafer. There was a variation of oxide thickness and hence, of the conducting properties from chip to chip, probably caused by temperature gradients during the sample fabrication (growth of the thermal oxide). The effects discussed below however were reproducibly observed on several samples with various oxide thickness.

{\bf The measurements} were carried out in a temperature range 2-20 K using PPMS-9 cryomagnetic system. The measurement current was in the range 50-200 nA to ensure absence of overheating. All measurements were performed in the frequency range 13-18Hz using lock-in amplifiers.

\begin{figure}
\vspace{0.1 in}
\centerline{\psfig{figure=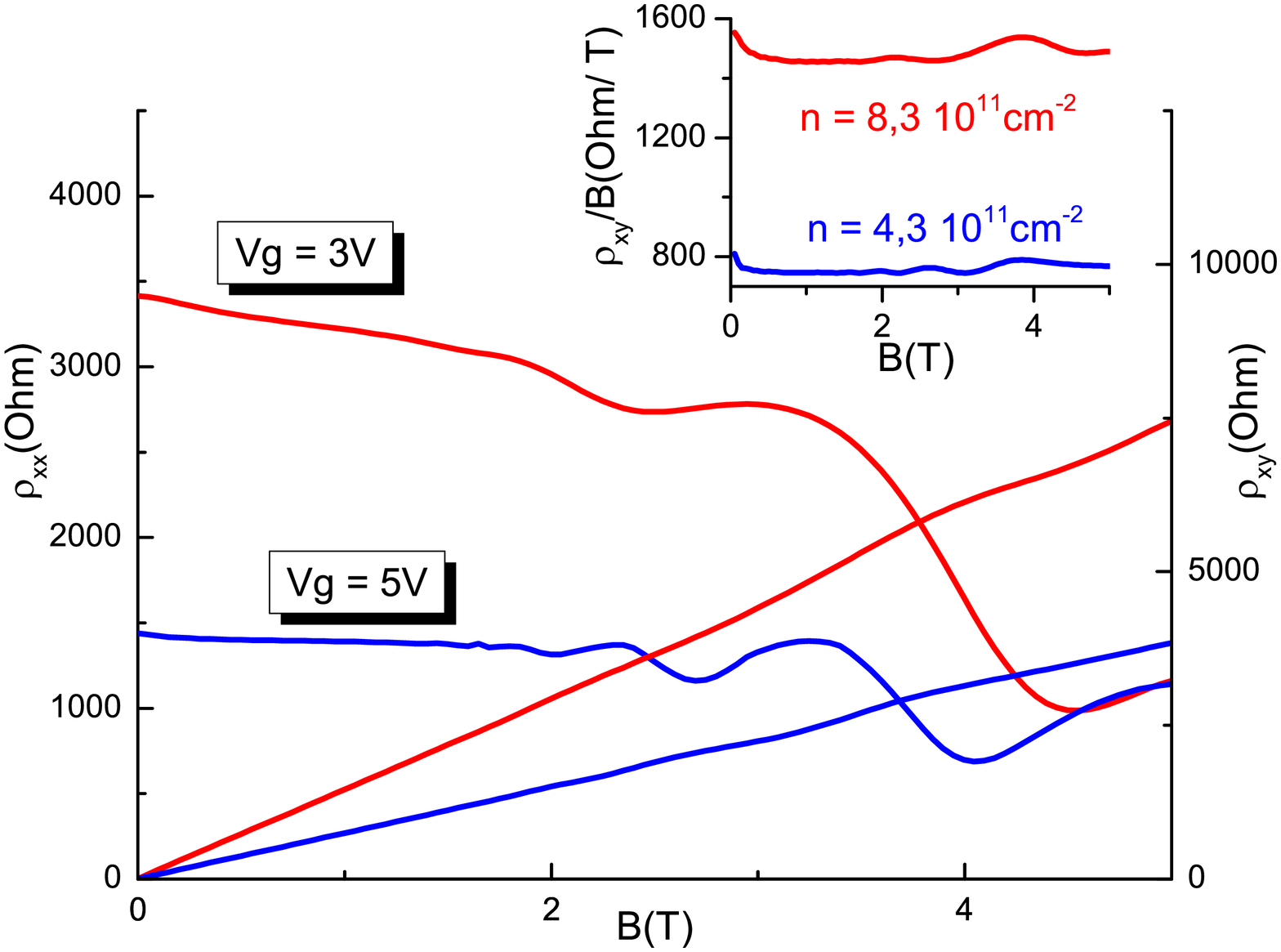,width=250pt}}
\begin{minipage}{3.2in}
\caption{ Magnetoresistance and Hall resistance Si-MOSFET of Hall bar geometry for two reperesentative gate voltages at $T=2$K. The insert shows Hall slope ($R_{xy}/B$) versus magnetic field.}
\label{figBare}
\end{minipage}
\vspace{0.4 in}
\end{figure}

{\bf Results:}
Bare 2D gas manifested mobility about 8000 cm$^2$/Vs at 4.2K. Magnetoresistance $\rho_{xx}(B)$ was weak, mainly due to quantum effects similar(weak localization, Shubnikov de Haas oscillations) to those observed previously in Si-MOSFETs (see Ref.\cite{ando} and references therein).  Hall resistance is a linear function of magnetic field, as it should be (See. Fig. \ref{figBare}).  In the lowest fields (as indicated in the inset to Fig.\ref{figBare}) there is a small feature close to $B=0$ probably related to weak localization but still not understood \cite{minkovHall, kuntsevichEEI}. The density of the electrons, extracted from Hall slope is roughly proportional to the gate voltage $n\propto V_g$. Resistivity of the bare 2DEG does not show up any hysteresis with gate.

\begin{figure}
\vspace{0.1 in}
\centerline{\psfig{figure=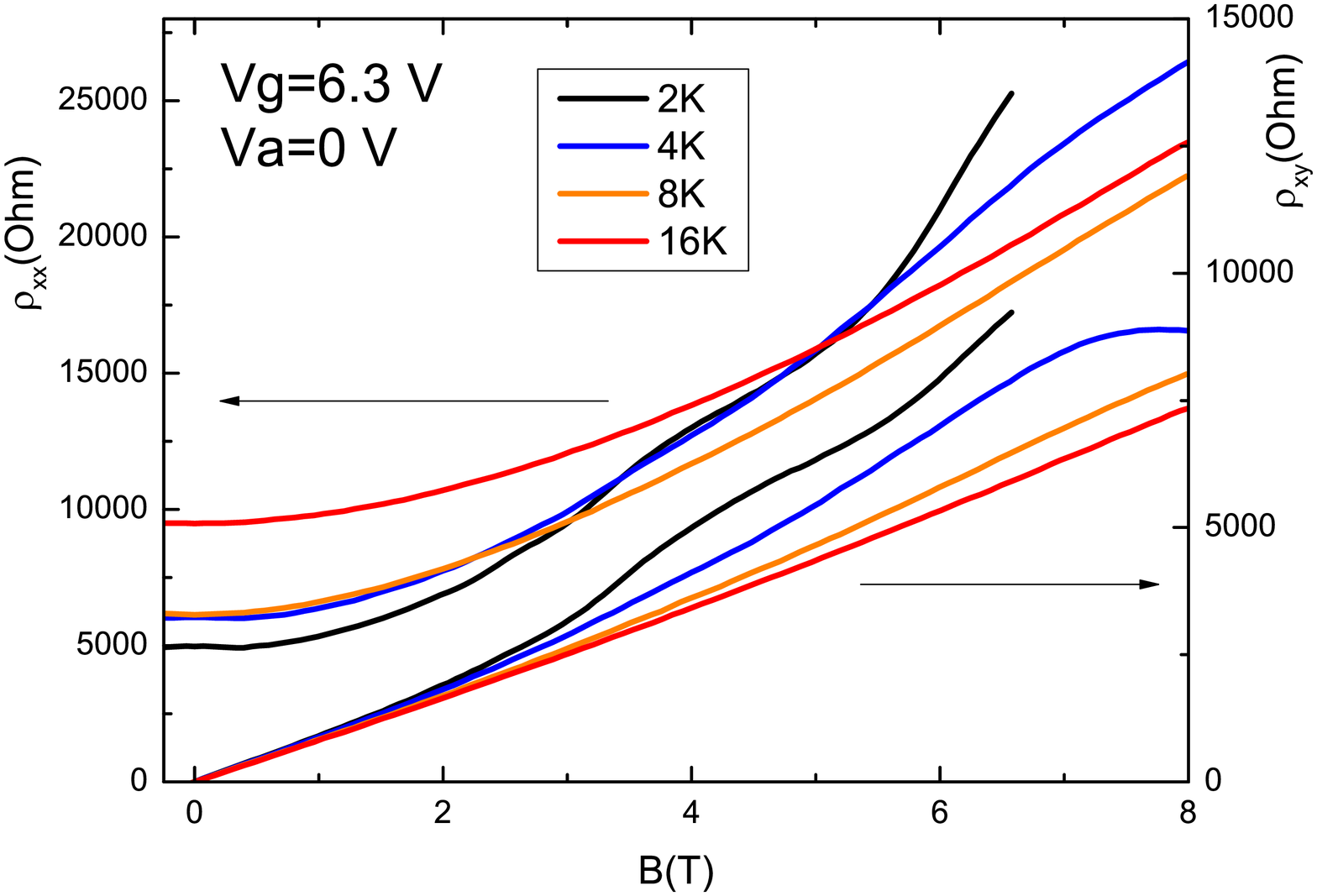,width=250pt}}
\begin{minipage}{3.2in}
\caption{Magnetoresistance and Hall resistance for R2DEG (sample AA1) $V_g=6.3V$, $V_a=0$V for various temperatures, indicated in the inset.}
\label{fig1}
\end{minipage}
\vspace{0.4 in}
\end{figure}

\begin{figure}
\vspace{0.1 in}
\centerline{\psfig{figure=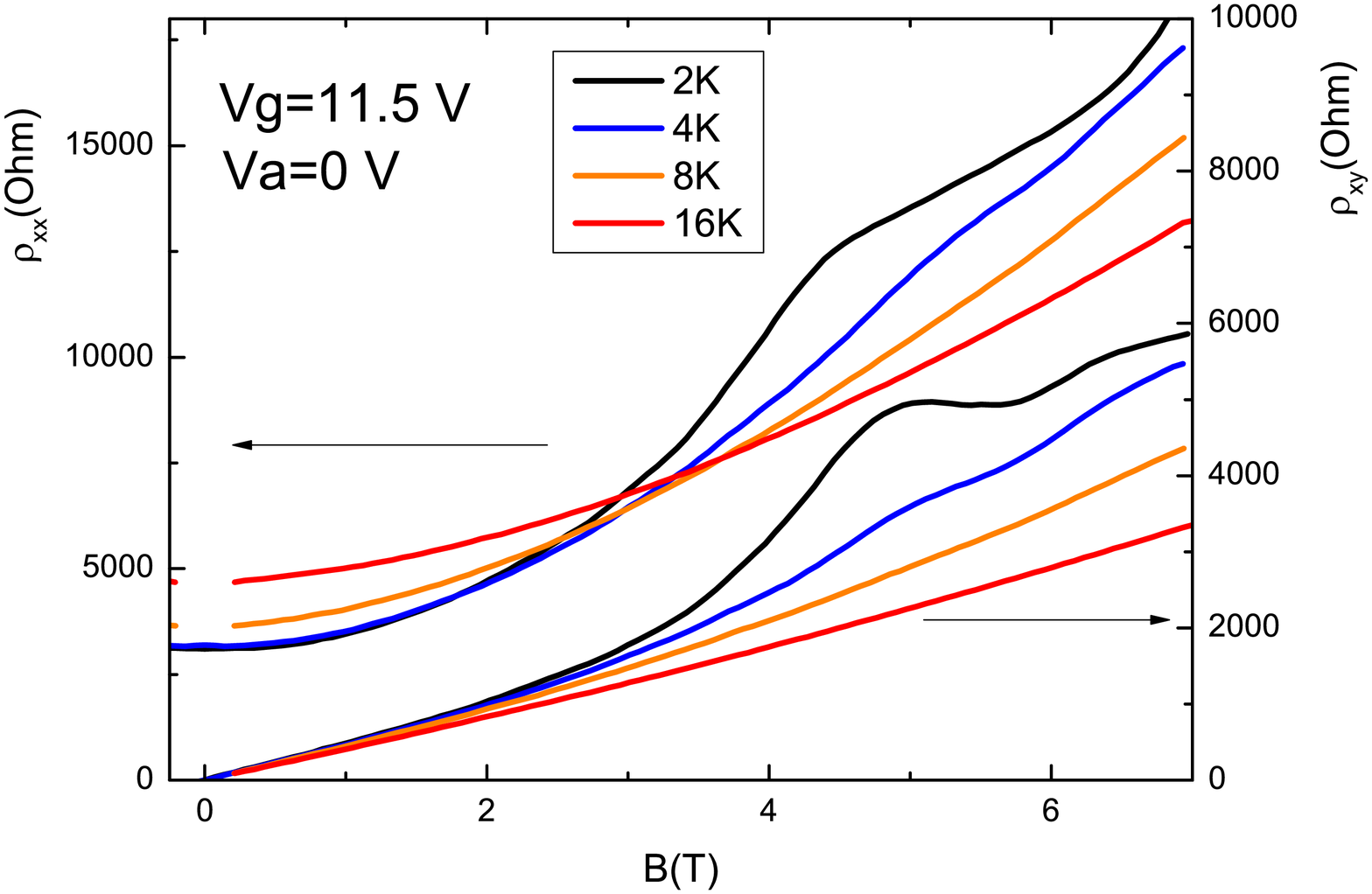,width=250pt}}
\begin{minipage}{3.2in}
\caption{Magnetoresistance and Hall resistance for R2DEG (sample AA1) $V_g=11.5V$, $V_a=0$V for various temperatures, indicated in the inset.}
\label{fig2}
\end{minipage}
\vspace{0.4 in}
\end{figure}

The resistivity of R2DEG is larger that for bare 2DEG(this is evident because of constricted geometry and effectively larger $l/w$ ratio).
The deviations of the {\it magneto}resistivity tensor for R2DEG from the that for the bare 2D gas become more and more pronounced as dots get more and more depleted. The $V_a$ value however should not be negative, in order to avoid irreversible changes due to recharging of impurity states. R2DEG demonstrated relaxation processes after changing the gate voltage. Equilibration time became larger with decreasing the temperature. In order to ensure equilibrium state we always swept gate voltages at elevated temperatures. We present below the most representatative figures for $Va=0$ in wide range of densities ($V_g$) (See Figs. \ref{fig1},\ref{fig2}).

With antidots the behavior of Hall effect and magnetoresistance changed considerably: magnetoresistance became essentially positive, Hall effect started to deviate from linear-in-B behavior. Positive magnetoresistance in non-uniform system is not surprising and follows from simple geometrical constriction \cite{litle2}, leading even to magnetosensors applications at room temperature\cite{solin} whereas the correction to Hall effect is much less trivial.

\begin{figure}
\vspace{0.1 in}
\centerline{\psfig{figure=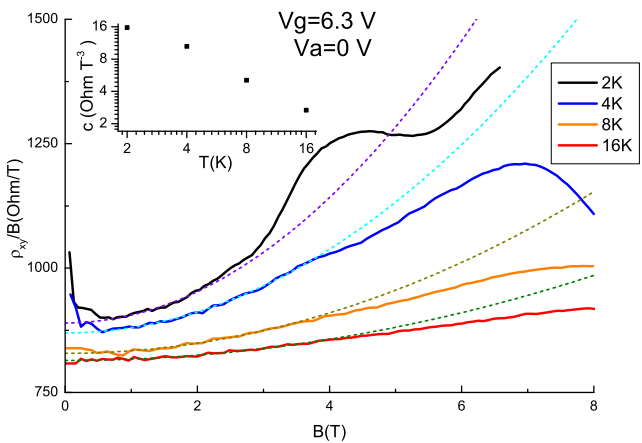,width=250pt}}
\begin{minipage}{3.2in}
\caption{Hall slope for R2DEG (sample AA1) $V_g=6.3V$, $V_a=0$V for various temperatures. Dashed lines - parabolic-in-field correction to Hall slope. Insert shows the temperature of the parabola coefficient.}
\label{fig3}
\end{minipage}
\vspace{0.4 in}
\end{figure}

Hall slope versus magnetic field is shown in Fig.\ref{fig3}. It differs dramatically from those in the inset to Fig. \ref{figBare}, the features of the Hall slope are as follows: (i) Similarly to the bare 2DEG the Hall effect has low-field feature and traces of Landau quantization; (ii) Zero-field Hall slope is temperature dependent, it looks like carriers are frozen out as $T$ decreases; (iii) The coefficient $c$ in low field expansion of the Hall slope ($\rho_{xy}/B\approx a + cB^2$) is temperature dependent and behaves approximately as $1/T$; (iv) The coefficient $c$ weakly depends on carrier density.

{\bf Discussion.}
The above observations imply that the AA gives rise to correction to the Hall resistance, similarly to additional scatterers add up to longitudinal resistance. This correction seems to be density independent in a certain density range and sensitive to temperature and magnetic field.

An explanation of the phenomenon is required. The essential ingredients of such explanation are as follows:
\begin{itemize}
\item{Some figures: The diameter of the antidot is about 3.5 um, the distance between antidots is 1.5 um or effectively less(see AFM image in Fig. \ref{sample}). Typical density of the 2DEG $5 - 10 \cdot 10^{11}$cm$^{-2}$. Mean free path $\sim 80$ nm.}
\item{The antidots themselves are non-conductive ($V_a=0$), however they are surrounded with weakly conductive shell with decreased electron density. Typical width of the shell might be about 100-200 nm. The width is determined by  gate oxide thickness.}
\item{The regime of nonlinear Hall effect falls in the crossover from ordinary classical behavior ($\mu B>1$) to quantization ($\hbar\omega_c>T$,$\Gamma$)}
\item{For Si-MOSFETs of similar mobility ($\sim 1$m$^2$/Vs) resistivity has strong "metallic" temperature dependence in the range of densities $n\sim2-5\times10^{11}$ cm${-2}$ and temperatures $T~2-10$K}
\end{itemize}

\begin{figure}
\vspace{0.1 in}
\centerline{\psfig{figure=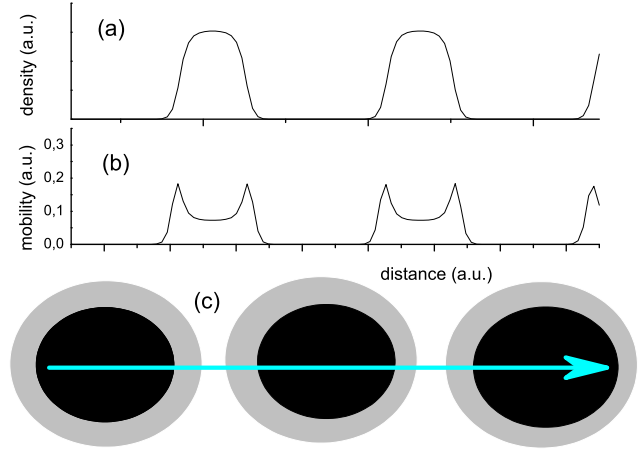,width=250pt}}
\begin{minipage}{3.2in}
\caption{Hall slope and magnetoresistance for sample AA2. (a) Carrier density distribution in lateral direction. (b) Mobility distribution. (c) Schematics of the antidots with shell.}
\label{expl}
\end{minipage}
\vspace{0.4 in}
\end{figure}

On the basis of the above facts we suggest the {\bf following possible explanation} of the phenomenon:
an interplay between bulk and interface scatterers leads to nonmonotonic density dependence of mobility in Si-MOSFETS\cite{ando}. In 2DEG with AA (see Fig. \ref{expl}c) density distribution is shown schematically in Fig. \ref{expl}a, while the mobility distribution is show in \ref{expl}b.
In magnetic field, high-mobility region, i.e. shells, contributes to Hall effect and makes Hall resistance larger because density in the shells is lower than in R2DEG. The mobility of the carriers in the shell behaves as inverse resistivity of the Si-MOSFET in the domain of the metallic conductivity ($1/\sigma\propto R\propto A+ B\cdot T$, see \cite{pudballistic}). This dependence leads to $\sim 1/T$ behavior of the correction to the Hall coefficient. At higher temperatures the mobility peak becomes shallower and correction to Hall effect disappears.

\begin{figure}
\vspace{0.1 in}
\centerline{\psfig{figure=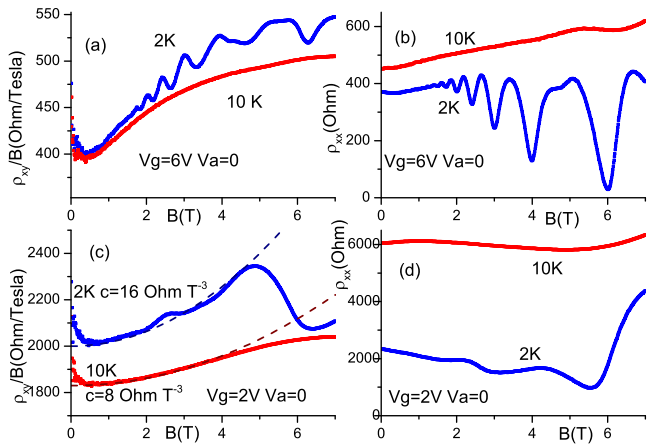,width=250pt}}
\begin{minipage}{3.2in}
\caption{Hall slope(panels a and c) and magnetoresistance (panels b and d) for sample AA2 at 2K (blue curves) and 10K (red curves). $Va=0$ everywhere.  $Vg=6$V (panels a and b). $Vg=2$V (panels c and d). In panel c dashed lines indicate parabolic-in-field correction to Hall slope. Constant $c$ of this parabola is indicated in the panel.}
\label{aa2}
\end{minipage}
\vspace{0.4 in}
\end{figure}

The lateral width of the shell is determined by geometry of the system (gate thickness). If the density, and hence conductivity of the 2D gas becomes very high, the size and contribution of high mobility low density area should become negligible. Fig. \ref{aa2}a,b demonstrates the effect of increased density we for the sample AA2, that has thinner oxide layer compared to AA1 and hence thinner shells of the antidots. It is easy to see that although Hall effect nonlinearity exists, its temperature dependence disappears, as expected. Because of large carrier density a whole palette of magnetooscillations from the bare 2D gas becomes clearly seen. We note that in sample AA2, in spite of such high electron densities ($\sim 1.5\cdot 10^{12}$ cm$^{-2}$), another fingerprint of nonuniform system, i.e. positive magnetoresistance, is observed, similarly to sample AA1.

When we decrease the density (see Fig. \ref{aa2}c,d), the behavior changes: magnetoresistance at low T becomes negative due to 2DEG, as it uses to be close to metal-insulator transition, and correction to the Hall coefficient becomes temperature dependent. The dependence is weaker than 1/T ($c=8$ Ohm$\cdot$T$^{-3}$ at 10 K and $c=16$ Ohm$\cdot$T$^{-3}$ at 2 K). We believe this is because 2DEG itself is closer to peak mobility and has its own strong $R(T)$ dependence, so contribution of the shells to the Hall resistance on top of the R2DEG is not as as distinguishable as in sample AA1. The strength of "metallicity" is also well seen from Fig.\ref{aa2}d: as T increases by factor of 5, resistivity triples.

To summarize, we have observed a novel effect in magnetotransport of the macroscopic gate-defined antidot array: positive and temperature dependent correction to the Hall resistance. We believe that this correction originates from the shells of the antidots, which have higher carrier mobility than the bare 2D gas. Our considerations however require further justification by microscopic theory.

{\bf Acknowledgments}. AYuK acknowledges support by RFBR (14-02-01169). We are thankful to V. Tripathi and V.M.Pudalov for discussions and reading the manuscript and Alexey Usoltsev for his assistance in the beginning of the measurements. Magnetotransport measurements have been done using research equipment of the LPI  Shared Facility Center.


\begin{thebibliography}{100}
\bibitem{solin}
Solin et al, Science {\bf 289}, 1530 (2000).
\bibitem{aas}
B.L. Altshuler,A.G. Aronov, B.Z. Spivak, Pis'ma Zh. Exp. Teor. Fiz {\bf 33}, 101 (1984) [JETP Lett. {\bf 33}, 94 (1984)]
\bibitem{weiss}
D. Weiss, M.L. Roukes, A. Menschig, P. Grambow, K. von Klitzing, Phys. Rev. Lett. {\bf 66}, 2790 (1991)
\bibitem{hofs}
C. R. Dean,	et al, Nature, {\bf 497}, 598–602 (2013); L.A. Ponomarenko et al, Nature {\bf 497}, 594-597 (2013);
\bibitem{isich} M. B. Isichenko, Rev. Mod. Phys. {\bf 64}, 961 (1992);
\bibitem{bulgad1} S.A. Bulgadaev, F.V. Kusmartsev, Phys. Lett. A {\bf 336}, 223-234 (2005)
\bibitem{bulgad2} S.A. Bulgadaev, F.V. Kusmartsev, Pis'ma v ZhETF, {\bf 81} (3), 157-161 (2005) [JETP Lett., {\bf 81}(3), 125-130 (2005)];
\bibitem{litle2}
M. M. Parish and P. B. Littlewood,  Nature (London) {\bf 426}, 162 (2003);
\bibitem{reznikov}
A.B. Berkut, Yu. V. Dubrovskii, M. S. Nunuparov, M. I. Reznikov, V. I Tal'yanskii, Pis'ma Zh. Exp. Teor. Fiz {\bf 44}, 254 (1986) [JETP Lett. {\bf 44}, 324 (1986)]
\bibitem{ando} T.~Ando, A.~B.~Fowler, and F.~Stern, Rev. Mod. Phys. {\bf 54}, 437, (1982).
\bibitem{minkovHall}
G. M. Minkov, A. V. Germanenko, O. E. Rut, A. A. Sherstobitov, and B. N. Zvonkov, Phys. Rev. B {\bf 82}, 035306 (2010).
\bibitem{kuntsevichEEI}
A.Yu. Kuntsevich, L.A. Morgun, V.M. Pudalov, Phys. Rev. B {\bf 87}, 205406 (2013);
\bibitem{pudballistic}
V. M. Pudalov, M. E. Gershenson, H. Kojima, G. Brunthaler, A. Prinz, and G. Bauer
Phys. Rev. Lett. {\bf 91}, 126403 (2003)
\end{thebibliography}
\end{document}